\documentclass[12pt]{article}

\def\UseBibLatex{1}
\def\BibLatexMode{1}

\makeatletter
\def\input@path{{styles/}}
\makeatother

\usepackage{latexml}

\usepackage{graphicx}
\usepackage{euscript}%
\usepackage{caption}
\usepackage{pifont}%
\usepackage{amsmath}%
\usepackage{amssymb}%
\usepackage{mathabx}%
\usepackage{xcolor}%
\usepackage{paralist}%
\usepackage{caption}%
\usepackage{xspace}%
\usepackage[cm]{fullpage}%
\usepackage{caption}%
\usepackage{mleftright}%

\usepackage[amsmath,thmmarks]{ntheorem}%
\theoremseparator{.}%

\usepackage{titlesec}
\titlelabel{\thetitle. }

\usepackage{hyperref}%
\hypersetup{%
   breaklinks,%
   colorlinks=true,%
   linkcolor=[rgb]{0.45,0.0,0.0},%
   citecolor=[rgb]{0,0,0.45}
}

\iflatexml
else
    \usepackage[ocgcolorlinks]{ocgx2}
\fi

\usepackage[hypcap=false]{caption}

\usepackage[bibencoding=utf8,style=alphabetic,backend=biber]{biblatex}%
\usepackage{sariel_biblatex}%

\newtheorem{theorem}{Theorem}[section]

\theoremseparator{}
\newtheorem{theoremf}[theorem]{Theorem}
\newtheorem{lemmaf}[theorem]{Lemma}
\theoremseparator{,}

\newtheorem{lemma}[theorem]{Lemma}

\theoremstyle{plain}%
\theoremheaderfont{\sf}%
\theorembodyfont{\upshape}%
\newtheorem*{remark:unnumbered}[FakeCounter]{Remark}%

\newtheorem*{defn:unnumbered}[FakeCounter]{Definition}
\newtheorem{defn}[theorem]{Definition}

\theoremheaderfont{\em}%
\theorembodyfont{\upshape}%
\theoremstyle{nonumberplain}%
\theoremseparator{}%

\theoremsymbol{}
\newtheorem{proofnoqed}{Proof:}

\theoremsymbol{\myqedsymbol}%
\newtheorem{proof}{Proof:}%

\theoremstyle{plain}%

\providecommand{\si}[1]{#1}

\newcommand{\myqedsymbol}{\rule{2mm}{2mm}}

\renewcommand{\Re}{\mathbb{R}}

\newcommand{\ts}{\hspace{0.6pt}}%
\newcommand{\MakeBig}{\rule[-.2cm]{0cm}{0.4cm}}
\newcommand{\MakesBig}{\rule[0.0cm]{0.0cm}{0.32cm}} %
\newcommand{\MakeSBig}{\rule[0.0cm]{0.0cm}{0.35cm}} %
\newcommand{\etal}{\textit{et~al.}\xspace}

\newcommand{\HLinkShort}[2]{\hyperref[#2]{#1\ref*{#2}}}
\newcommand{\HLink}[2]{\hyperref[#2]{#1~\ref*{#2}}}
\newcommand{\HLinkPage}[2]{\hyperref[#2]{#1~\ref*{#2}%
      $_\text{p\pageref{#2}}$}}
\newcommand{\HLinkPageOnly}[1]{\hyperref[#1]{Page~\refpage*{#1}%
      $_\text{p\pageref{#1}}$}}

\newcommand{\HLinkSuffix}[3]{\hyperref[#2]{#1\ref*{#2}{#3}}}
\newcommand{\HLinkPageSuffix}[3]{\hyperref[#2]{#1\ref*{#2}%
      #3$_\text{p\pageref{#2}}$}}

\newcommand{\apndlab}[1]{\label{apnd:#1}}
\newcommand{\apndref}[1]{\HLink{Appendix}{apnd:#1}}

\newcommand{\seclab}[1]{\label{sec:#1}}

\newcommand{\lemlab}[1]{\label{lemma:#1}}
\newcommand{\lemref}[1]{\HLink{Lemma}{lemma:#1}}%

\newcommand{\thmlab}[1]{{\label{theo:#1}}}
\newcommand{\thmref}[1]{\HLink{Theorem}{theo:#1}}

\providecommand{\deflab}[1]{\label{def:#1}}
\newcommand{\defref}[1]{\HLink{Definition}{def:#1}}

\newcommand{\figlab}[1]{\label{fig:#1}}
\newcommand{\figref}[1]{\HLink{Figure}{fig:#1}}

\newcommand{\atgen}{\symbol{'100}} \newcommand{\SarielThanks}[1]{%
   \thanks{Department of Computer Science; University of Illinois; 201
      N. Goodwin Avenue; Urbana, IL, 61801, USA; {\tt
         sariel\atgen{}illinois.edu}; {\tt
         \url{http://www.illinois.edu/\string~sariel/}.} #1}}

\newlength{\ppicwd}

\newcommand{\BallSet}{\mathcal{B}}%
\newcommand{\BallSetA}{\mathcal{C}}

\newcommand{\SepSet}{S}

\newcommand{\sphere}{\mathsf{s}}

\renewcommand{\th}{th\xspace}

\newcommand{\ball}{\mathsf{b}}%
\newcommand{\ballA}{\mathsf{b}'}

\newcommand{\DSet}{\EuScript{D}}%
\newcommand{\PntSet}{{P}}%
\newcommand{\disk}{\mathsf{d}}%
\newcommand{\diskA}{\mathsf{f}}%
\newcommand{\diskB}{\mathsf{u}}%
\newcommand{\diskC}{\mathsf{v}}%
\newcommand{\pnt}{\mathsf{p}}%
\newcommand{\pntA}{\mathsf{q}}%
\newcommand{\pntB}{\mathsf{t}}%
\newcommand{\AreaX}[1]{\mathrm{area}\pth{#1}}

\newcommand{\G}{{G}}%

\newcommand{\Vertices}{{V}}%
\newcommand{\Edges}{{E}}%

\newcommand{\Vd}{v_d}%
\newcommand{\constDbl}{\ell_d}%
\newcommand{\constDblTwo}{\ell_2}%

\newcommand{\VolX}[1]{\mathrm{vol}\pth{#1}}
\newcommand{\ballX}[1]{\mathrm{ball}\pth{#1}}

\newcommand{\Holder}{H\"older\xspace}

\newcommand{\distX}[2]{\left\| {#1} - {#2} \right\|}%

\newcommand{\pth}[1]{\mleft({#1}\mright)}

\newcommand{\Cone}{\psi}%
\definecolor{blue25}{rgb}{0,0,0.55}%
\newcommand{\emphic}[2]{%
   \textcolor{blue25}{%
      \textbf{\emph{#1}}}%
   \index{#2}}

\newcommand{\emphi}[1]{\emphic{#1}{#1}}

\newcommand{\Ex}[2][\!]{\mathop{\mathbf{E}}#1\pbrcx{#2}}
\newcommand{\Prob}[1]{\mathop{\mathbf{Pr}}\!\pbrcx{#1}}
\newcommand{\cardin}[1]{\left| {#1} \right|}%
\newcommand{\norm}[1]{\left\| {#1} \right\|}
\newcommand{\pbrcx}[1]{\left[ {#1} \right]}

\newcommand{\DSetSmall}{\DSet_{\leq \ell}}
\newcommand{\DSetLarge}{\DSet_{> \ell}}

\newcommand{\diskY}[2]{\mathrm{disk} \pth{#1, #2}}
\newcommand{\rad}{\rho}%
\newcommand{\areaX}[1]{\mathrm{area}\pth{#1}}%

\newlength{\savedparindent}
\newcommand{\SaveIndent}{\setlength{\savedparindent}{\parindent}}
\newcommand{\RestoreIndent}{\setlength{\parindent}{\savedparindent}}

\numberwithin{figure}{section}%
\numberwithin{table}{section}%
\numberwithin{equation}{section}%

\newlength{\CharHeight}
\AtBeginDocument{\setlength{\CharHeight}{\fontcharht\font`X}}   %

\IfFileExists{figs/qed_duck.pdf}%
{%
   \renewcommand{\myqedsymbol}{%
      \reflectbox{\includegraphics[height=1.4\CharHeight]%
         {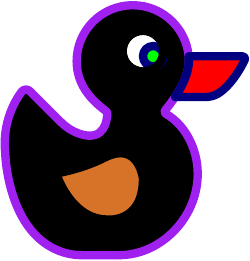}}%
   }%
}{}

\providecommand{\TPDF}[2]{\texorpdfstring{#1}{#2}}

\bibliography{separator}

\begin{document}

\title{A Simple Proof of the Existence of a Planar Separator}

\author{%
   Sariel Har-Peled%
   \SarielThanks{Work on this paper was partially supported by NSF AF awards CCF-1421231, CCF-1217462, and CCF-0915984.  }%
}

\date{April 30, 2013\footnote{\LaTeX{}ed on \today.}}

\maketitle

\begin{abstract}
    We provide a simple proof of the existence of a planar separator by showing that it is an easy consequence of the circle packing theorem. We also reprove other results on separators, including: \smallskip
    \begin{compactenum}[\qquad(A)]
        \item There is a simple cycle separator if the planar graph is triangulated. Furthermore, if each face has at most $d$ edges on its boundary, there is a cycle separator of size $O\bigl( \sqrt{ d n} \,\bigr)$.

        \smallskip%
        \item For a set of $n$ balls in $\Re^d$, that are $k$-ply, there is a separator, in the intersection graph of the balls, of size $O\pth{ k^{1/d}n^{1-1/d}}$.

        \smallskip%
        \item The $k$ nearest neighbor graph of a set of $n$ points in $\Re^d$ contains a separator of size $O\pth{ k^{1/d}n^{1-1/d}}$.
    \end{compactenum}
    \smallskip%
    The new proofs are (arguably) significantly\footnote{Or insignificantly, or not at all. I am willing to support all sides of this argument. The skeptical reader can replace the above sentence by ''The new proofs are newer than the older proofs.''} simpler than previous proofs.
\end{abstract}

\section{Introduction}

The \emphi{planar separator theorem} is a fundamental result about planar graphs \cite{u-tpg-51,lt-stpg-79}. Informally, it states that one can remove $O\pth{\! \sqrt{n}\ts \MakesBig }$ vertices from a planar graph with $n$ vertices and break it into ``significantly'' smaller parts. It is widely used in algorithms to facilitate efficient divide-and-conquer schemes on planar graphs. For further details on planar separators and their applications, see Wikipedia (\url{http://en.wikipedia.org/wiki/Planar_separator_theorem}).

Here, we present a simple proof of the planar separator theorem. Most of the main ingredients of the proof are present in earlier work on this problem; see Miller \etal \cite{mttv-sspnng-97}, Smith and Wormald \cite{sw-gsta-98}, and Chan \cite{c-ptasp-03}. Furthermore, the constants in the separator we get are inferior to known constructions \cite{ast-ps-94}. See \thmref{separator} for the exact statement.

Nevertheless, the new proof is relatively self-contained and (arguably) simpler than previous proofs. We also reprove some of the other results of Miller \etal \cite{mttv-sspnng-97} and Miller \cite{m-fsscs-86}. Again, our proofs are arguably simpler (but the constants are inferior).

\SaveIndent%

\section{Proof of the planar separator theorem}

\subsection{The proof}
\seclab{proof}%

Given a planar graph $\G=(V, E)$ it is known that it can be drawn in the plane as a \emphi{kissing graph}; that is, every vertex is a disk, and an edge in $\G$ implies that the two corresponding disks touch (this is known as Koebe's theorem or the cycle packing theorem, see \cite{pa-cg-95}). Furthermore, all these disks are interior disjoint.

Let $\DSet$ be the set of disks realizing $\G$ as a kissing graph, and let $\PntSet$ be the set of centers of these disks. Let $\disk$ be the smallest radius disk containing $n/10$ of the points of $\PntSet$, where $n= \cardin{\PntSet} = \cardin{V}$. To simplify the exposition, we assume that $\disk$ is of radius $1$ and is centered at the origin. Randomly pick a number $x \in [1,2]$ and consider the circle $C_x$ of radius $x$ centered at the origin. Let $S$ be the set of all disks in $\DSet$ that intersect $C_x$. We claim that, in expectation, $S$ is a good separator.

\begin{lemma}
    \lemlab{separates}%
    The separator $S$ breaks $\G$ into two subgraphs with at most $(9/10) n$ vertices in each connected component.
\end{lemma}

\begin{proofnoqed}
    The circle $C_x$ breaks the graph into two components: (i) the disks with centers inside $C_x$, and (ii) the disks with centers outside $C_x$.

    \medskip%

    \noindent%
    \begin{minipage}{0.79\linewidth}
        \RestoreIndent{}%

        The corresponding vertices in $\G$ are disconnected once we remove $S$. Furthermore, a disk of radius $2$ can be covered by $9$ disks of radius $1$, as depicted in \figref{double_cover}. As such, the disk of radius $2$ at the origin can contain at most $9n/10$ points of $\PntSet$ inside it, as a disk of radius $1$ can contain at most $n/10$ points of $\PntSet$. We conclude that there are at least $n/10$ disks of $\DSet$ with their centers outside $C_x$, and, by construction, there are at least $n/10$ disks of $\DSet$ with centers inside $C_x$. Once $S$ is removed, no connected component of the graph $\G \setminus S$ can be larger than $(9/10)n$.
        \hfill \myqedsymbol
    \end{minipage}
    \hfill%
    \begin{minipage}{0.2\linewidth}
        \vspace{-0.5cm}%
        \hfill%
        \includegraphics{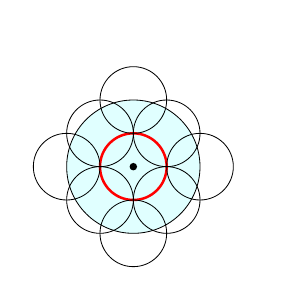}
        \hfill%
        \captionof{figure}{}
        \figlab{double_cover}
    \end{minipage}
\end{proofnoqed}
\smallskip

\begin{lemma}
    \lemlab{small}%
    We have $\Ex{\MakeSBig\! \cardin{S}} \leq 11 \sqrt{n}$, where $n = \cardin{V}$.
\end{lemma}

\begin{proof}
    Let $\ell \in (0,1)$ be a parameter to be specified shortly. We split $\DSet$ into two sets: $\DSetSmall$ and $\DSetLarge$ of all disks of diameter $\leq \ell$ and $>\ell$, respectively.

    \smallskip
    \noindent%
    \begin{minipage}{0.84\linewidth}
        \RestoreIndent{}%
        Consider the ring $R = \diskY{0}{x+\ell} \,\setminus \, \diskY{0}{x-\ell}$, and observe that any disk $\diskA$ of $\DSetLarge$ that intersects $C_x$, must contain inside it a disk of radius $\ell/2$ that is fully contained in $R$. As such, $\diskA$ covers an area of size at least $\alpha = \pi (\ell/2)^2$ of this ring. The area of $R$ is
        \begin{math}
            \beta = \pi \pth{\pth{x+\ell}^2 - \pth{x-\ell}^2} = 4\pi x \ell.
        \end{math}
        As such, the number of disks of $\DSetLarge$ that intersect $C_x$ is
        \begin{math}
            \leq \beta/\alpha = 4\pi x \ell/(\pi \ell^2/4) = 16x /\ell.
        \end{math}
        As $\Ex{x}=3/2$, we have $\Ex{\beta/\alpha} = 24/\ell$.
    \end{minipage}%
    \begin{minipage}{0.15\linewidth}
        \vspace{-0.5cm}
        \hfill {\includegraphics{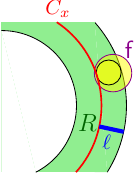}}
    \end{minipage}

    \smallskip%
    Consider a disk $\diskB_i \in \DSetSmall$ of radius $r_i$ centered at $\pnt_i$.  The circle $C_x$ intersects $\diskB_i$ if and only if $x \in [\norm{\pnt_i} - r_i, \norm{\pnt_i} + r_i]$, and as $x$ is being picked uniformly from $[1,2]$, the probability for that is at most $2r_i/|2-1| = 2r_i \leq \ell$.  Since $\cardin{\DSetSmall} \leq n$, we have that the expected number of disks of $\DSetSmall$ that intersect $C_x$ is at most $n \ell$. Adding the two quantities together, we have that the expected number of disks intersecting $C_x$ is bounded by $n \ell + 24/\ell$, which is $\leq 2\sqrt{24 n}$, for $\ell = 1/\sqrt{24 n}$.
\end{proof}

Now, putting \lemref{separates} and \lemref{small} together implies the following.

\begin{theorem}%
    \thmlab{separator}%
    Let $\G = (\Vertices,\Edges)$ be a planar graph with $n$ vertices. There exists a set $S$ of at most $11 \sqrt{n}$ vertices of $\G$, such that removing $S$ from $\G$ breaks it into several connected components, each containing at most $(9/10)n$ vertices.
\end{theorem}%

\paragraph{Remarks.}
    \begin{figure}[t]%
    \centerline{%
       \begin{tabular}{ccccc}
         \includegraphics[page=1]{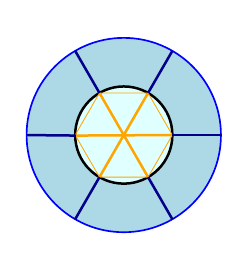}%
         &\qquad&
                  \includegraphics[page=2]{figs/hex_2}%
         &\qquad\quad&
                       \includegraphics[page=3]{figs/hex_2}\\
         (A) && (B) && (C)
       \end{tabular}%
    }
    \caption{How to cover a disk of radius $2$ by $7$ disks of radius
       $1$.}
    \figlab{hex}
\end{figure}%
(A) The constant in \lemref{small} can be improved by working (a bit) harder and using the Cauchy-Schwarz inequality. For completeness, we provide the proof in \apndref{better:c}.

(B) The main difference between the proof of \thmref{separator} and the work of Miller \etal \cite{mttv-sspnng-97} is that they found the cycle $C_x$ by lifting the disks to a sphere in 3d, using conformal mapping to recenter the resulting caps on the sphere around the center point of the centers of the caps. Our direct packing argument avoids these stages. We also avoid using the Cauchy-Schwarz inequality.

(C) As suggested by G\"u\si{nter} Rote, one can improve the constant of \thmref{separator} to $7/8$ (instead of $9/10$) by using a tiling that uses only $7$ disks instead of $9$, see \figref{hex}. It is easy to verify that $7$ disks are needed for such a cover.

\section{Extensions}

\subsection{Weighted version}

\begin{lemma}%
    \lemlab{separator:w}%
    Let $\G = (\Vertices,\Edges)$ be a planar graph with $n$ vertices, and assume that the vertices have non-negative weights assigned to them, with total weight $W$. There exists a set $S$ of $4 \sqrt{n}$ vertices of $\G$, such that removing $S$ from $\G$ breaks it into several connected components, each containing a set of vertices of total weight at most $(9/10)W$.
\end{lemma}

\begin{proof}
    The proof of \thmref{separator} goes through, with the minor modification that $\disk$ is picked to be the smallest disk, such that the total weight of the centers of the disks it covers is $\geq W/10$.
\end{proof}

Note that if there is a vertex in the graph with weight $\geq W/10$, then the returned separator could be this single vertex, which is a legal answer (as the weight of the remaining graph is sufficiently small).

\subsection{Cycle separators}

A planar graph $\G$ is \emphi{maximal} if one can not add edges without violating its planarity. Any drawing of a maximal planar graph is a triangulation. Namely, every face is a triangle. But then, in the realization of the graph as a kissing graph of disks, a face of the complement of the union of the disks has three touching disks as its boundary.

\medskip
\noindent%
\begin{minipage}{0.75\linewidth}
    \RestoreIndent%

    In particular, consider the separating cycle $C_k$, and two disks $\diskA$ and $\diskA'$ that intersect it consecutively along $C_x$. Let $I$ be interval on $C_x$ between $\diskA \cap C_x$ and $\diskA'\cap C_x$. The interval $I$ belongs to a single face of the complement of the union of disks, and in particular, this face has both $\diskA$ and $\diskA'$ on its boundary. As such, the vertices of $\G$ that correspond to $\diskA$ and $\diskA'$ are connected by an edge. That is, the resulting separator is a cycle in $\G$. This cycle is simple since $C_x$ intersects a disk along an interval (or not at all). Thus, we get the following.
\end{minipage}
\hfill%
\begin{minipage}{0.24\linewidth}
    \hfill%
    \includegraphics{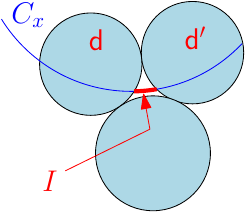}
\end{minipage}

\begin{theoremf}\emph{\textbf{\cite{m-fsscs-86}.}}
    \thmlab{separator:2}%
    Let $\G = (\Vertices,\Edges)$ be a maximal planar graph with $n$ vertices. There exists a set $\SepSet$ of $4 \sqrt{n}$ vertices of $\G$, such that removing $S$ from $\G$ breaks it into several connected components, each containing at most $(9/10)n$ vertices. Furthermore, $\SepSet$ is a simple cycle in $\G$.
\end{theoremf}

\subsubsection{Cycle separator if the graph is not triangulated.}

\begin{lemmaf}\emph{\textbf{\cite{m-fsscs-86}.}}
    \lemlab{separator:2:x}%
    Let $\G = (\Vertices,\Edges)$ be a connected planar graph with $n$ vertices, where the $i$\th face has $d_i$ vertices on its boundary, and let $N = \sum_{i} d_i^2$. Then, there exists a set $\SepSet$ of $4 \sqrt{N}$ vertices of $\G$, such that removing $S$ from $\G$ breaks it into several connected components, each containing at most $(9/10)n$ vertices. Furthermore, $\SepSet$ is a cycle in $\G$.

    In particular, if the maximum face degree in $\G$ is $d$, then the separator size is $O\pth{\!\sqrt{nd}}$.
\end{lemmaf}
\begin{proof}
    The idea is to fill in the faces of $\G$ so they are all triangulated.

    So, consider a cycle $C$ (not necessarily simple -- an edge might be traversed twice) with $k$ vertices that forms the boundary of a single face in the given embedding of $\G$. Next, we build a graph having $C_1= C$ as its outer boundary, as follows -- it has $k$ copies of $C$ one inside the other, where the $i$\th copy $C_i$ is connected to the $i-1$ and $i+1$ copies, in the natural way, where a vertex is connected to its copies. Drawn in the plane, this results in a grid-like construction. We also arbitrarily triangulate the innermost copy $C_k$, and every quadrilateral face is triangulated arbitrarily. The resulting graph $\G_C$ has $k^2$ vertices, and has the property that any path between any two vertices of $C$ in $\G_C$, the corresponding shortest path in $C$ is shorter (or of the same length). See \figref{fill_in} for an example.

    \smallskip%
    \noindent%
    \begin{minipage}{0.7\linewidth}
        \RestoreIndent%
        We repeat this fill-in process for all the faces of $\G$, and let $\G'$ be the resulting graph. $\G'$ is still planar, and the number of resulting vertices in the new graph is $N= \sum_{i} d_i^2$. Observe that $\sum_i d_i \leq 6n$, as every vertex $v$ incident on a face $r$, can be charged to an edge adjacent to $v$ and $f$. If done consistently, an edge would be charged at most twice, and the maximum number of edges in a planar graph is $3n -6$ by Euler's formula.

        In particular, if the maximum value of $d_i$ is $d$, then the maximum of $N = \sum_i d_i^2$ is $O(nd)$, as can be easily verified.
    \end{minipage}%
    \begin{minipage}{0.29\linewidth}
        \vspace{-0.3cm}%
        \hfill%
        \includegraphics[scale=0.9]{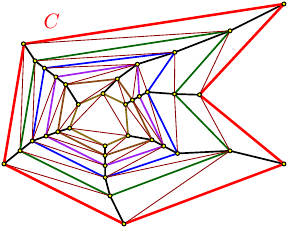}

        \vspace{-0.2cm}
        \captionof{figure}{}
        \figlab{fill_in}
    \end{minipage}%

    \smallskip%
    Now, we assign weight zero to all the newly introduced vertices in $\G'$ and weight one to the original vertices (that appear in $\G$). The graph $\G'$ is a fully triangulated planar graph with $N$ vertices. By \lemref{separator:w}, a separator provides the desired partition, and the number of vertices on this separator is $\leq 4 \sqrt{N}$. Since $\G'$ is triangulated, the separator is a simple cycle in $\G'$. We now replace portions of it that use the face grids with the appropriate paths along the original boundary of the faces. The resulting cycle in $\G$ has the same number of vertices, providing the same quality of separation (or better, since some vertices migrated to the separator), as desired.
\end{proof}

Miller's result is somewhat stronger than \lemref{separator:2:x}, as he assumes the graph is $2$-connected, and can ensure that in this case the separator is a \emph{simple} cycle.

\subsection{Ball systems that are \TPDF{$k$}{k}-ply}

A set of balls $\BallSet$ in $\Re^d$ is \emphi{$k$-ply}, if no point of $\Re^d$ is contained in more than $k$ balls of $\BallSet$.

\begin{defn}
    \deflab{doubling:constant}%
    The \emphi{doubling constant} of a metric space is the smallest number of balls of the same radius needed to cover a ball of twice the radius (formally, we take the maximum such number over all possible balls to be covered). The doubling constant of $\Re^d$ is $\constDbl \leq 2^{O(d)}$ \cite{v-cbseb-05}.
\end{defn}

\begin{theoremf}\emph{\textbf{\cite{mttv-sspnng-97}.}}
    \thmlab{sep_k_ply}%
    Let $\BallSet$ be a set of $n$ balls that is $k$-ply in $\Re^d$. Then, there exists a sphere $\sphere$ that intersects $4 k^{1/d} n^{1-1/d}$ balls of $\BallSet$. Furthermore, the number of balls of $\BallSet$ that are completely inside (resp. outside) $\sphere$ is $\geq n/(\constDbl+1)$.
\end{theoremf}

\begin{proof}
    Let $\PntSet$ be the set of centers of the balls of $\BallSet$. As above, let $\ball$ be the smallest ball containing $n/(1+\constDbl)$ points of $\PntSet$. As above, assume that $\ball$ is centered at the origin and has radius $1$. Let $\sphere$ be a random sphere centered at the origin with radius $x$ picked randomly from the range $[1,2]$.

    Now, arguing as above, there are at most $\pth{\constDbl /(\constDbl + 1)}n $ points of $\PntSet$ inside $\sphere$, and as such, at least $(1- \constDbl/(\constDbl + 1 )) = n/(\constDbl+1)$ points of $\PntSet$ outside $\sphere$. As such, $\sphere$ is a good separator for the balls.

    As for the expected number of balls intersecting $\sphere$, let $\Vd r^d$ be the volume of a ball of radius $r$ in $\Re^d$, where $\Vd$ is a constant that depends on the dimension. As above, we clip the balls of $\BallSet$ to the ball of radius $2$ centered at the origin, replacing every lens by an appropriate ball of the same volume. Let $\rad_i$ denote the radius of the $i$\th such ball $\ballA_i$, for $i=1,\ldots, n$. By the $k$-ply property, we have that
    \begin{align*}
        \sum_i \rad_i^d%
        =%
        \frac{1}{\Vd} \pth{\sum_i \Vd \rad_i^d}%
        \leq%
        \frac{k}{\Vd} \VolX{\MakeSBig \ballX{2}}%
        \leq%
        k 2^d,
    \end{align*}
    where $\ballX{2}$ denotes a ball of radius $2$ in $\Re^d$. As before, the probability of the $i$\th ball to intersect $\sphere$ is bounded by $2\rad_i$. Let $\SepSet$ be the set of balls of $\BallSet$ that intersect $\sphere$. We have, by \Holder's inequality, that
    \begin{align*}
        \Ex{\cardin{\SepSet} \MakeBig\! }%
        &=%
        \sum_i \Prob{ \ballA_i \cap \sphere \ne \emptyset \MakeBig }%
        \leq%
        \sum_i 2\rad_i = 2 \sum_i 1 \cdot \rad_i%
        \leq%
        2 \pth{\sum_{i=1}^n 1^{d/(d-1)}}^{(d-1)/d} \pth{\sum_{i=1}^n
           \rad_i^d}^{1/d}%
        \\ &\leq%
        2 n^{1-1/d} \pth{k 2^d }^{1/d} \leq 4 n^{1-1/d} k^{1/d},
    \end{align*}
    as desired.
\end{proof}

\subsection{Separators for the \TPDF{$k$}{k}\th nearest neighbor %
   graph}

Let $\PntSet$ be a set of $n$ points in $\Re^d$, and let $k$ be a parameter. The \emphi{$k$\th nearest neighbor graph} $\G_k = (\PntSet, \Edges)$ is the graph, where two points $\pnt, \pntA \in \PntSet$ are connected by an edge $\pnt \pntA \in \Edges$, if $\pntA$ is the $i$\th nearest neighbor of $\pnt$ in $\PntSet$ (or $\pnt$ is the $i$\th nearest neighbor of $\pntA$), for $i \leq k$.

\begin{theoremf}\emph{\textbf{\cite{mttv-sspnng-97}.}}
    Let $\PntSet$ be a set of $n$ points in $\Re^d$, and let $k$ be a parameter. The $k$\th nearest neighbor graph $\G_k = (\PntSet, \Edges)$ has a separator of size $O(k^{1/d} n^{1-1/d})$, such that each connected component has at most $\pth{\constDbl/(\constDbl +1)} n$ vertices, where $\constDbl$ is the doubling constant of $\Re^d$, see \defref{doubling:constant}.
\end{theoremf}

\begin{proof}
    We follow the proof of Miller \etal \cite{mttv-sspnng-97}. A point $\pntA \in \PntSet$ is an \emphi{$i$-client} of $\pnt \in \PntSet$, if $\pnt$ is the $i$\th nearest neighbor of $\pntA$, for $i\leq k$.  If $\pntA$ is a $k$-client of $\pnt$, then create a ball of radius $\distX{\pnt}{\pntA}$ centered at $\pntA$.  Let $\BallSet$ be the resulting set of $n$ balls. The key observation is that this set of balls is $O(k)$-ply -- which we prove here using a standard argument.

    We claim that every point $\pnt \in \PntSet$ can serve at most $O(k)$ clients. To this end, cover the sphere of directions around $\pnt$ with cones with angular diameter at most $30^\degree$. It is easy to verify that $c = 2^{O(d-1)}$ cones are needed at most.

    \medskip%

    \noindent%
    \begin{minipage}{0.76\linewidth}
        \RestoreIndent%
        The key observation is now that for any two points $\pntA,\pntB \in \PntSet$ that belong to the same cone $\Cone$ of $\pnt$, it must be that $\distX{\pntA}{\pntB} \leq \distX{\pnt}{\pntB}$, assuming that $\pntA$ is closer to $\pnt$ than $\pntB$, as an easy geometric argument shows. That is, if $\pntA_1,\ldots, \pntA_k$ are the $k$ closest points to $\pnt$ in $\PntSet \cap \Cone$, then these are the only points of $\PntSet \cap \Cone$ that might be $k$-clients of $\pnt$. It follows that $\pnt$ can have at most $c k$ $k$-clients, and its degree in $\G_k$ is $\leq c k + k$. The maximum degree of a vertex in $\G_k$ is $O(k)$.
    \end{minipage}
    \hfill%
    \begin{minipage}{0.21\linewidth}
        \hfill \includegraphics{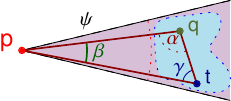}
    \end{minipage}

    \smallskip%
    To see why this implies that the set of balls $\BallSet$ is $k$-ply, consider any point $\pnt \in \Re^d$, insert it into $\PntSet$, and observe that the degree of $\pnt$ in the graph $\G_{k+1}$ bounds the number of balls of $\BallSet$ that cover it. By the above, this is $O(k)$, as desired.

    By \thmref{sep_k_ply}, there are $4 k^{1/d} n^{1-1/d}$ balls of $\BallSet$, such that their removal breaks the intersection graph of $\BallSet$ into connected components each of size at most $\pth{ \constDbl/(\constDbl +1)} n$. The corresponding set of points of $\PntSet$ is the desired separator of $\G_k$.
\end{proof}

\subsection{Separator for \TPDF{$r$}{r} vertices in a planar graph}

Our purpose here is to show that in a triangulated planar graph, there is always a cycle of size $O(\sqrt{r})$ whose removal separates (roughly) $r$ vertices from the remainder of the graph. To this end, we need the following.

\begin{lemma}%
    \lemlab{packing:r}%
    Let $\BallSet$ be a set of $n$ balls in $\Re^d$ that are interior disjoint, and let $r> 0$ be some prespecified integer number. Let $\ball$ be the smallest ball that contains $r$ centers of the balls of $\BallSet$. Then $\ball$ intersects at most $\pth{\constDbl}^2 \pth{r + 1} $ balls of $\BallSet$. Furthermore, $2\ball$ intersects at most $\pth{\constDbl}^3 \pth{r + 1} $ balls of $\BallSet$, where $\constDbl$ is the doubling constant of $\Re^d$, see \defref{doubling:constant}.
\end{lemma}
\begin{proof}
    Assume $\ball$ is of radius one and centered at the origin. Consider the ball $4\ball$, and observe that it can be covered by $\pth{\constDbl}^2$ balls of radius one, and let $\BallSetA$ be this set of balls. As such, $4\ball$ contains at most $\pth{\constDbl}^2 r$ centers of balls of $\BallSet$. Any other ball of $\BallSet$ that intersects $\ball$ must have a radius of at least $3$, as its center is at a distance of at least $4$ from the origin.

    It is easy to verify that such a ball $\ball'$ must contain fully at least one ball of $\BallSetA$. Indeed, consider the segment connecting the center of $\ball'$ with the origin, and consider the point on this segment on $\partial 4\ball$. Clearly, this point must be covered by one of the balls of $\BallSetA$, and this ball is fully contained in $\ball'$.
\end{proof}

\begin{lemma}
    Let $\G$ be a planar graph with $n$ vertices, and let $r > 0$ be a sufficiently large integer. There exists a set of vertices $\SepSet$ of size $\leq 4 \constDblTwo \sqrt{r}$, such that $\G \setminus \SepSet$ is disconnected into two sets of vertices, $X$ and $Y$, such that $r/2\constDblTwo \leq \cardin{X} \leq r$, where $\constDblTwo$ is a constant (see \defref{doubling:constant}).  Furthermore, if $\G$ is triangulated, then $\SepSet$ is a cycle in the graph.
\end{lemma}

\begin{proof}
    Let $\BallSet$ be the realization of $\G$ as a kissing graph of interior disjoint disks. Let $\disk$ be the smallest disk containing $r/\constDblTwo$ centers of $\BallSet$, and assume it is of radius one and centered at the origin.  \lemref{packing:r} implies that $2\disk$ intersects at most $r\pth{\constDblTwo}^2$ disks of $\BallSet$, and let $\BallSetA$ be this set of balls. Now consider the circle $C_x$ centered at the origin of radius $x$, where $x$ is picked randomly and uniformly from the range $[1,2]$. Let $\SepSet$ be the set of disks of $\BallSetA$ that intersect $C_x$.

    Now, by the analysis of \lemref{small}, the expected number of disks of $\BallSetA$, and thus of $\BallSet$ that intersects $C_x$ is $\leq 4 \sqrt{ \cardin{\BallSetA}} \leq 4 \constDblTwo\sqrt{r}$.  This implies that the number of disks strictly inside $C_x$ is at least $r/\constDblTwo - 4 \constDblTwo\sqrt{r} \geq r/2\constDblTwo$, if $r \geq 64 \pth{\constDblTwo}^4$. Similarly, it is easy to argue that $C_x$ contains at most $r$ disks of $\BallSet$.
\end{proof}

\section{Conclusions}

This write-up demonstrates that the planar separator theorem is an easy consequence of the circle packing theorem, originally proved by Paul Koebe in 1936 \cite{k-kdka-36}. The circle packing theorem is thus the ``true'' magic -- converting a topological property (a graph being planar) into a packing property (i.e., disks touching each other).

\paragraph{An open problem.}

The current algorithmic proofs of the circle packing theorem build an evolving discrete structure that keeps improving after each iteration, till in the limit it converges to the desired packing. Specifically, no finite algorithm computes the realization of a planar graph as a circle packing.

It seems unlikely that a finite algorithm is possible because of numerical issues. However, a much weaker version is sufficient for the planar separator theorem. In particular, can one find a set of disks for a planar graph, such that two vertices are connected if their respective disks intersect (in their interiors), and no point in the plane is contained in more than, say, $c$ disks of this set, where $c$ is some universal constant (thus, we allow disks to intersect even if their corresponding vertices are not connected in the planar graph).

\paragraph{Acknowledgments}

The author thanks Mark \si{de} Berg, Timothy Chan, Robert Krauthgamer, G\"u\si{nter} Rote, and Christian Sommer for useful comments on the manuscript. The idea of using a ring area argument, in the proof of \lemref{small}, came about during discussions with Mark \si{de} Berg. G\"u\si{nter} Rote suggested the elegant tilling depicted in \figref{hex}.

\printbibliography

\appendix

\section{Proof of \TPDF{\lemref{small}}{Lemma 2.2} with a better %
   constant}
\apndlab{better:c}

\begin{proof}
    Consider a disk $\diskB_i$ of $\DSet$ of radius $r_i$ centered at $\pnt_i$.  If $\diskB_i$ is fully contained in $\diskA_2$ (the disk of radius $2$ centered at the origin), then the circle $C_x$ intersects $\diskB_i$ if and only if $x \in [\norm{\pnt_i} - r_i, \norm{\pnt_i} + r_i]$, and as $x$ is being picked uniformly from $[1,2]$, the probability for that is at most $2r_i/|2-1| = 2r_i$. In this case, we set $\rad_i = r_i$ and $\diskC_i = \diskB_i$ for reasons that would become clear shortly.

    \medskip%

    \noindent%
    \begin{minipage}{0.77\linewidth}
        \RestoreIndent{}%
        Otherwise, if $\diskB_i$ is not fully contained in $\diskA_2$ then the set $L_i = \diskB_i \cap \diskA_2$ is a ``lens''. Consider a disk $\diskC_i$ of the same area as $L_i$ contained inside $\diskA_2$ and tangent to its boundary. Clearly, if $C_x$ intersects $\diskB_i$ then it also intersects $\diskC_i$, see figure on the right. Furthermore, the radius of $\diskC_i$ is $\rad_i = \sqrt{\areaX{ \diskB_i \cap \diskA_2} /\pi}$, and, by the above, the probability that $C_x$ intersects $\diskC_i$ (and thus $\diskB_i$) is at most $2\rad_i$.

        Observe that as the disks of $\DSet$ are interior disjoint, we have that $\sum_i \rad_i^2 = \sum_i \areaX{\diskB_i \cap \diskA_2}/\pi \leq \AreaX{\diskA_2}/\pi = 4$.  Now, by linearity of expectation and the Cauchy-Schwarz inequality, we have that
    \end{minipage}
    \begin{minipage}{0.22\linewidth}
        \hfill%
        \includegraphics{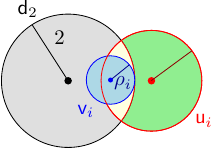}
    \end{minipage}

    \begin{align*}%
        \Ex{\cardin{S} \MakeBig\! }%
        &=%
        \Ex{\MakeBig\! \cardin{ \DSet \cap C_x}}%
        =%
        \sum_i \Prob{ \diskB_i \cap C_x \ne \emptyset \MakeBig }%
        \leq%
        \sum_i \Prob{ \diskC_i \cap C_x \ne \emptyset \MakeBig }%
        \leq%
        \sum_i 2\rad_i = 2 \sum_i 1 \cdot \rad_i%
        \\%
        & \leq%
        2 \sqrt{\sum_{i=1}^n 1^2} \sqrt{\sum_{i=1}^n \rad_i^2}%
        \leq%
        2 \sqrt{n} \sqrt{4} = 4 \sqrt{n}.
    \end{align*}
\end{proof}

\end{document}